\documentclass[
  sigconf,
  screen,
]{acmart}

\AtBeginDocument{%
  }

\copyrightyear{2026}
\acmYear{2026}
\setcopyright{cc}
\setcctype{by}
\acmConference[UMAP '26]{34th ACM Conference on User Modeling, Adaptation and Personalization}{June 08--11, 2026}{Gothenburg, Sweden}
\acmBooktitle{34th ACM Conference on User Modeling, Adaptation and Personalization (UMAP '26), June 08--11, 2026, Gothenburg, Sweden}
\acmDOI{10.1145/3774935.3806172}
\acmISBN{979-8-4007-2311-7/2026/06}

\usepackage{enumitem}

\setlist{nosep,leftmargin=*,itemsep=2pt,topsep=3pt}

\begin{document}

\title{Can Third-Party Annotators Reliably Evaluate Conversational Recommender Systems?}

\author{Michael Müller}
\orcid{0009-0005-0918-8833}
\affiliation{%
  \institution{Department of Computer Science, University of Innsbruck}
  \city{Innsbruck}\country{Austria}}
\email{michael.m.mueller@uibk.ac.at}

\author{Amir Reza Mohammadi}
\orcid{0000-0003-3934-6941}
\affiliation{%
  \institution{Department of Computer Science, University of Innsbruck}
  \city{Innsbruck}\country{Austria}}
\email{amir.reza@uibk.ac.at}

\author{Andreas Peintner}
\orcid{0000-0001-7337-524X}
\affiliation{%
  \institution{Department of Computer Science, University of Innsbruck}
  \city{Innsbruck}\country{Austria}}
\email{a.peintner@gmx.net}

\author{Beatriz Barroso Gstrein}
\orcid{0009-0006-9111-5402}
\affiliation{%
  \institution{Department of Computer Science, University of Innsbruck}
  \city{Innsbruck}\country{Austria}}
\email{beatriz.barroso@uibk.ac.at}

\author{Günther Specht}
\orcid{0000-0003-0978-7201}
\affiliation{%
  \institution{Department of Computer Science, University of Innsbruck}
  \city{Innsbruck}\country{Austria}}
\email{guenther.specht@uibk.ac.at}

\author{Eva Zangerle}
\orcid{0000-0003-3195-8273}
\affiliation{%
  \institution{Department of Computer Science, University of Innsbruck}
  \city{Innsbruck}\country{Austria}}
\email{eva.zangerle@uibk.ac.at}

\renewcommand{\shortauthors}{M. Müller et al.}

\begin{abstract}
User-centric evaluation has become a key paradigm for assessing Conversational Recommender Systems (CRS), aiming to capture subjective qualities such as satisfaction, trust, and rapport. To enable scalable evaluation, recent work increasingly relies on third-party annotations of static dialogue logs by crowd workers or large language models. However, the reliability of this practice remains largely unexamined.  
In this paper, we present a large-scale empirical study investigating the reliability and structure of user-centric CRS evaluation on static dialogue transcripts. We collected 1,053 annotations from 124 crowd workers on 200 ReDial dialogues using the 18-dimensional CRS-Que framework. Using random-effects reliability models and correlation analysis, we quantify the stability of individual dimensions and their interdependencies.  
Our results show that utilitarian and outcome-oriented dimensions such as accuracy, usefulness, and satisfaction achieve moderate reliability under aggregation, whereas socially grounded constructs such as humanness and rapport are substantially less reliable. Furthermore, many dimensions collapse into a single global quality signal, revealing a strong halo effect in third-party judgments.  
These findings challenge the validity of single-annotator and LLM-based evaluation protocols and motivate the need for multi-rater aggregation and dimension reduction in offline CRS evaluation.
\end{abstract}



\maketitle

\section{Introduction and Related Work}

Conversational Recommender Systems (CRS) have become a vital component of modern online services, enabling natural multi-turn interactions through which users can express preferences and receive personalized recommendations and explanations~\cite{Jannach2021}.
Recent advances in Large Language Models (LLMs) have further improved the fluency and contextual understanding of such systems~\cite{Zhang2024, Wang2024a}, yet their evaluation remains a persistent challenge~\cite{Jannach2023}.

Traditional evaluation approaches rely primarily on system-centric offline metrics, such as Recall@K for recommendation accuracy or BLEU for response generation quality.
However, these metrics fail to capture essential aspects of the user experience (UX), including empathy, rapport, and adaptability~\cite{Jin2024, Manzoor2024, Mueller2025}.
Recent work has shown that static accuracy metrics correlate poorly with self-reported user satisfaction~\cite{Bernard2025b} and that reference-based metrics often do not reflect human preferences~\cite{Gienapp2025}.

To address this gap, the community increasingly adopts user-centric evaluation frameworks~\cite{Bauer2025}.
Instruments such as ResQue~\cite{Pu2011} and its CRS-specific extension CRS-Que~\cite{Jin2024} operationalize subjective qualities of interaction into multi-dimensional questionnaires.
While interactive user studies provide reliable insights into these dimensions~\cite{Yun2025}, they are costly and difficult to scale.
As a result, many studies rely on \emph{third-party evaluation}, where crowd workers or LLMs assess static dialogue logs (e.g., from benchmarks such as ReDial~\cite{Li2018}) as proxies for the original user experience.

This shift toward scalable evaluation has accelerated the development of LLM-based evaluators such as CoRE~\cite{Chen2025b} and CONCEPT~\cite{Huang2024}.
Although these approaches include limited human validation, they do not systematically examine inter-rater reliability or the validity of individual dimensions.
More broadly, a central assumption remains largely untested: that an external observer can reliably infer subjective user perceptions from static dialogue transcripts alone.

Current evaluation protocols rely on two implicit premises.
First, user-centric dimensions are assumed to be measurable in a stable and reproducible way from static dialogue logs.
Second, user-centric frameworks such as CRS-Que treat these dimensions as distinct constructs, although it is unclear whether external annotators meaningfully differentiate between them.
Together, these concerns motivate a systematic analysis of both the reliability of individual dimensions and their empirical interdependencies in static evaluation settings.
We therefore address the following research questions:

\begin{itemize}[leftmargin=*]
    \item \textbf{RQ1 (Reliability):} To what extent can individual \mbox{CRS-Que} dimensions be reliably assessed by crowd workers on static dialogue logs?
    \item \textbf{RQ2 (Structure):} How are these dimensions interrelated, and do they form distinct clusters or a single global quality signal?
\end{itemize}

In this paper, we present a large-scale empirical study of crowdsourced user-centric evaluation of CRS.
We collect annotations from 124 crowd workers on 200 ReDial dialogues using the 18-dimensional \mbox{CRS-Que} framework and analyze them with random-effects reliability models and correlation analysis.
Our results reveal a clear dichotomy.
Annotators can form consistent judgments about utilitarian and outcome-oriented qualities such as accuracy and intention to use, but struggle to reliably assess socially grounded constructs such as humanness and rapport.
Moreover, judgments tend to collapse into a single global quality impression, producing a pronounced halo effect across dimensions.
To support reproducibility, we provide our annotation data and analysis code at \url{https://github.com/michael-mue/reliable-crs-eval}.

\section{Methodology}

This section details the experimental design, crowdsourcing procedure, and statistical methods used to assess the reliability of third-party CRS evaluation.

\subsection{Study Design}

\subsubsection{Dataset and Sampling} 
We utilized the ReDial~\cite{Li2018} dataset, which contains over 10k human human movie recommendation dialogues. We sampled 200 dialogues (preprocessed to expand movie entities) and assigned them to a pool of annotators such that every dialogue received at least five independent human ratings, resulting in a total of 1,053 evaluations.

\subsubsection{Task Procedure and Participants} 
We recruited participants via Prolific to evaluate the dialogues using the CRS-Que~\cite{Jin2024} Framework to evaluate the quality of conversational recommendations from a user-centric perspective.
Each participant was assigned a batch of 10 dialogues: 9 randomly sampled from the target pool and one ``quasi-gold'' dialogue (with known low quality) for validation.
For each dialogue, participants rated 18 items corresponding to CRS-Que dimensions (e.g., \emph{accuracy}, \emph{novelty}, \emph{rapport}) on a 5-point Likert scale (1=\emph{strongly disagree}, 5=\emph{strongly agree}). Annotators were instructed to adopt the perspective of the information seeker.
We recruited 124 participants with native English proficiency between December 2025 and January 2026. Participants were compensated £3.75 per submission (median completion time $\approx$ 23 minutes), corresponding to an hourly rate of over\ £9/hr.

\subsubsection{Quality Control} 
To mitigate satisficing and task fatigue, we implemented a strict quality assurance procedure: (1) \emph{Attention Checks}: Two explicit instruction-based checks were embedded as additional dimensions labeled ``Attention Check,'' prompting raters to select \emph{Strongly Agree} or \emph{Disagree};
(2) \emph{Quasi-Gold Dialogues}: Participants who failed to distinguish objectively nonsensical or incoherent control dialogues were removed; and (3) \emph{Response Patterns}: Submissions with unrealistic completion times were removed. After applying these filters, we retained 117 high-quality submissions (7 removed), providing adequate coverage for the 200 target dialogues, each dialogue receiving 4--8 ratings (on avg. 5.27).

\subsection{Statistical Framework}

To rigorously assess the quality and structure of the collected annotations, we apply a combination of power analysis, random-effects modeling, and rank-based agreement metrics.

\subsubsection{Power Analysis}

\begin{figure}
  \includegraphics[width=0.45\textwidth, trim=0cm 0.3cm 0cm 0.3cm, clip]{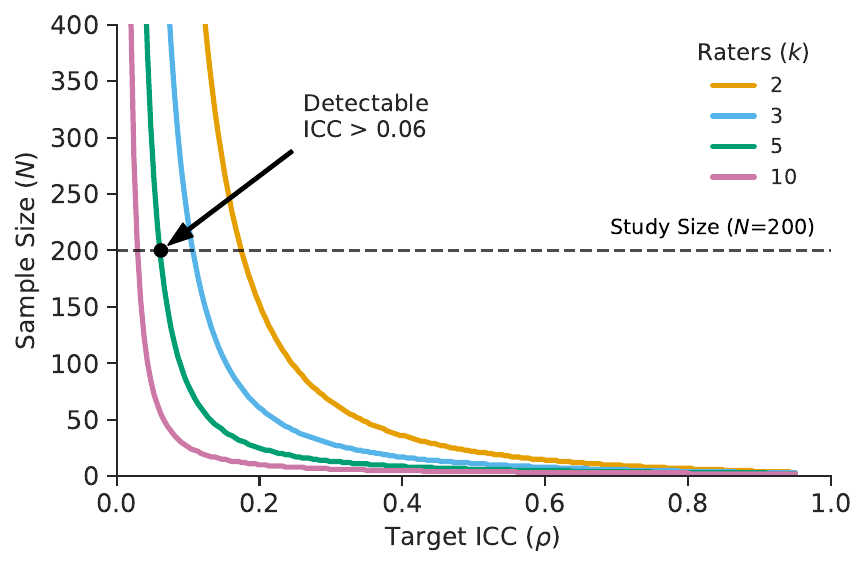}
  \Description[ICC precision plot]{ICC(2,5) precision plot showing confidence half-width versus sample size for inter-rater reliability.}
  \caption{Power Analysis for Inter-Rater Reliability}\label{fig:minimal_icc}
\end{figure}

To ensure our study is robust and cost-effective, we conducted an \textit{a priori} power analysis following the approximation method of Walter et al.~\cite{Walter1998}.
Our design goal was two-fold: (1) to achieve sufficient statistical power to detect non-random agreement ($H_0: \text{ICC} = 0$), and (2) to ensure precise estimation of reliability coefficients with narrow confidence intervals (RQ1).
As shown in Figure~\ref{fig:minimal_icc}, with $k=5$ raters per dialogue, a sample size of $N=200$ (dashed line) provides 80\% power ($\alpha=.05$) to detect an ICC as low as $\approx 0.06$.
This high sensitivity is critical for identifying dimensions where inter-rater agreement is present but weak, preventing Type II errors in the reliability analysis.
Regarding precision, assuming a moderate true reliability ($\rho=0.6$), the expected width of the 95\% confidence interval is approximately $0.12$ ($\pm 0.06$).
This level of precision ensures that observed reliability scores can be meaningfully categorized (e.g., distinguishing \textit{fair} from \textit{moderate} agreement).
Additionally, for the structural analysis (RQ2), a sample size of $N=200$ dialogues provides over 99\% power to detect medium correlations ($r=0.3$) and 80\% power to detect weak correlations ($r \approx 0.2$) between dimensions, ensuring a robust basis for the cluster analysis.

\subsubsection{Reliability}
To evaluate reliability, we calculate complementary agreement- and variance-based metrics that assess signal presence and stability under sparse crowdsourced annotation.
These measures characterize whether dialogue-level signals are detectable and how they behave under aggregation.

\textit{One-way random-effects model.}
We first report the Intra\-class Correlation Coefficient (ICC)~\cite{Koo2016, Shrout1979} based on a one-way random-effects model.
Ratings were modeled as
\begin{equation}
y_{ij} = \mu + u_i + \varepsilon_{ij},
\end{equation}
where $y_{ij}$ denotes the rating of annotator $j$ for dialogue $i$, $\mu$ is the
global mean, $u_i \sim \mathcal{N}(0, \sigma^2_{\text{dial}})$ represents the
random effect of dialogue $i$, and $\varepsilon_{ij} \sim
\mathcal{N}(0, \sigma^2_{\text{resid}})$ captures residual variance.
Based on these estimates, we computed
\begin{equation}
ICC(1) = \frac{\sigma^2_{\text{dial}}}
{\sigma^2_{\text{dial}} + \sigma^2_{\text{resid}}},
\end{equation}
quantifying the reliability of a \emph{single} randomly selected annotator
under strict absolute agreement assumptions. A zero ICC(1) thus reflects differing rater baselines (e.g., severity vs. leniency), not necessarily an absence of underlying signal.

\textit{Crossed random-effects model.}
To \looseness=-1 explicitly account for systematic annotator-specific scale effects~\cite{Artstein2008, McGraw1996}, we additionally fit a crossed random-effects model with random intercepts for both
dialogue and annotator:
\begin{equation}
y_{ij} = \mu + u_i^{(\text{dial})} + v_j^{(\text{rater})} + \varepsilon_{ij},
\end{equation}
where $u_i^{(\text{dial})} \sim \mathcal{N}(0, \sigma^2_{\text{dial}})$ captures
dialogue-level variance, while ${v_j^{(\text{rater})} \sim \mathcal{N}(0, \sigma^2_{\text{rater}})}$ captures systematic rater effects, and
$\varepsilon_{ij}$ denotes residual noise.
From this model, we report
\begin{equation}
  \text{Rel}_{\text{dial}}^{\text{single}} =
\frac{\sigma^2_{\text{dial}}}
{\sigma^2_{\text{dial}} + \sigma^2_{\text{rater}} + \sigma^2_{\text{resid}}},
\end{equation}
indicating whether a dialogue-level signal is present beyond rater
heterogeneity, without assuming aggregation.

\textit{Reliability of aggregated scores.}
As \looseness=-1 dialogue-level evaluation often relies on multiple annotations per
dialogue, we estimate the reliability of the \emph{mean} rating across $k$ annotators for both models.
For the strict one-way model (ICC(1,$k$)), the residual variance term $\sigma^2_{\text{resid}}$ is divided by $k$.
For the crossed model (Rel$_{\text{dial}}(k)$), the rater and residual variance components ($\sigma^2_{\text{rater}}$ and $\sigma^2_{\text{resid}}$) are divided by $k$.
We used the harmonic mean number of the raters per dialogue ($k \approx 5.14$).
For both models, the reliability of aggregated dialogue scores increases monotonically with the number of raters due to the averaging of variance components in the denominator (residual variance for ICC, and both rater and residual variance for the crossed model). However, gains diminish rapidly and are bounded by the amount of dialogue-level signal present, such that adding raters cannot compensate for inherently low signal-to-noise dimensions.

\textit{Rank-based agreement: Krippendorff’s $\alpha$ (ordinal).}
To assess agreement in relative judgments independent of absolute scale
alignment, we compute Krippendorff’s $\alpha$ using an ordinal distance metric
($\alpha_{\text{ord}}$).
Krippendorff’s $\alpha$ measures the extent to which annotators agree beyond
chance on the \emph{ordering} of dialogues and is robust to missing data and
unbalanced annotation designs, complementing the absolute and variance-based
reliability measures above.

\textit{Interpretation Guidelines.}
To interpret variance-based reliability coefficients (both ICC and Rel), we follow Koo and Li~\cite{Koo2016}: values less than 0.50 indicate poor reliability, between 0.50 and 0.75 moderate, and above 0.75 good reliability.
For Krippendorff’s $\alpha$, while standard guidelines suggest values $>0.67$ for tentative conclusions~\cite{Krippendorff2011}, we treat it as a descriptive indicator, acknowledging that subjective crowdsourcing often yields lower agreement due to genuine perceptual differences rather than error.

\section{Results}

We now present the empirical findings regarding the reliability of crowdsourced annotations (RQ1) and the latent structure of the evaluation dimensions (RQ2).

\subsection{RQ1: Reliability Analysis}
\begin{table}
\caption{Inter-Rater Reliability Metrics by Dimension.}
\label{tab:irr_summary}
\footnotesize
\begin{tabular}{lccccc}
\toprule
\textbf{Dimension} & \textbf{Rel$_{\text{dial}}^{(k)}$} & \textbf{Rel$_{\text{dial}}^{\text{single}}$} & \textbf{$\alpha_{\text{ord}}$} & \textbf{ICC(1)} & \textbf{ICC(1,$k$)} \\
\midrule
Accuracy & 0.69 & 0.30 & 0.28 & 0.00 & 0.00 \\
Satisfaction & 0.69 & 0.30 & 0.28 & 0.00 & 0.00 \\
Perceived Usefulness & 0.67 & 0.28 & 0.27 & 0.00 & 0.00 \\
CUI Understanding & 0.65 & 0.27 & 0.25 & 0.00 & 0.00 \\
Perceived Ease Of Use & 0.64 & 0.26 & 0.23 & 0.00 & 0.00 \\
Trust Confidence & 0.64 & 0.26 & 0.24 & 0.00 & 0.00 \\
CUI Attentiveness & 0.63 & 0.25 & 0.23 & 0.00 & 0.00 \\
CUI Adaptability & 0.62 & 0.24 & 0.24 & 0.00 & 0.00 \\
Intention To Use & 0.60 & 0.23 & 0.22 & 0.00 & 0.00 \\
Transparency & 0.60 & 0.22 & 0.22 & 0.22 & 0.60 \\
Intention To Purchase & 0.60 & 0.22 & 0.21 & 0.00 & 0.00 \\
CUI Response Quality & 0.59 & 0.22 & 0.21 & 0.00 & 0.00 \\
User Control & 0.59 & 0.22 & 0.21 & 0.00 & 0.00 \\
Novelty & 0.59 & 0.22 & 0.22 & 0.00 & 0.00 \\
Explainability & 0.55 & 0.19 & 0.19 & 0.00 & 0.00 \\
Interaction Adequacy & 0.48 & 0.15 & 0.14 & 0.00 & 0.00 \\
CUI Rapport & 0.47 & 0.15 & 0.15 & 0.00 & 0.00 \\
CUI Humanness & 0.41 & 0.12 & 0.14 & 0.00 & 0.00 \\
\bottomrule
\end{tabular}%
\end{table}

Table~\ref{tab:irr_summary} reports inter-rater reliability estimates for all CRS-Que dimensions using complementary reliability metrics.
Dimensions are ordered by aggregated dialogue-level reliability $\mathrm{Rel}_{\text{dial}}^{(k)}$, which reflects the stability of dialogue-level signals under aggregation.
Under the one-way random-effects model, single-rater reliability ICC(1) collapses to zero for the majority of dimensions, and aggregated reliability ICC(1,$k$) remains close to zero in most cases.
This indicates that, when rater-specific effects are not modeled explicitly, between-dialogue variance is largely dominated by residual variance.
In practice, this implies that systematic differences in rater baselines---where some annotators are consistently stricter than others---overshadow the actual variance in dialogue quality.
Consequently, metrics based on exact agreement underestimate reliability, failing to capture that raters agree on the relative ranking of the dialogues.
If annotations were merely statistical noise, the crossed model would also yield near-zero reliability. Instead, filtering out personal biases reveals a consistent dialogue-level signal across all dimensions, independently corroborated by our rank-based agreement ($\alpha_{\text{ord}}$).
Single-rating dialogue reliability $\mathrm{Rel}_{\text{dial}}^{\text{single}}$ ranges from 0.12 (CUI Humanness) to 0.30 (Accuracy and Satisfaction), demonstrating that individual annotations still contain measurable dialogue-specific information once systematic rater differences are accounted for.
When considering hypothetical aggregation, $\mathrm{Rel}_{\text{dial}}^{(k)}$ increases substantially for many dimensions, exceeding 0.6 for task- and outcome-oriented constructs such as Accuracy, Satisfaction, and Perceived Usefulness.
These results demonstrate that a meaningful dialogue signal exists, but it only becomes visible when the model explicitly controls for rater bias.
Krippendorff’s ordinal alpha $\alpha_{\text{ord}}$ exhibits moderate values across most dimensions, suggesting that annotators tend to agree on relative ordering even where absolute agreement is weak.
Notably, $\alpha_{\text{ord}}$ closely tracks the single-rater reliability estimates ($\mathrm{Rel}_{\text{dial}}^{\text{single}}$).
This convergence confirms that the primary source of disagreement in the unadjusted ICC(1) is systematic rater offset (severity/leniency), which both the rank-based and crossed-variance models effectively discount.
For Transparency, one-way and crossed reliability estimates coincide.
In this case, rater-specific variance is negligible, such that treating rater effects as residual noise does not distort dialogue-level variance estimates.
Taken together, these results indicate that low or zero one-way ICC values should not be interpreted as an absence of dialogue-level signal.
Instead, they reflect the limitations of strict absolute-agreement models in sparse crowdsourcing settings.
These findings suggest that relying on single annotations for ReDial, particularly for social dimensions, may be insufficient for reliable evaluation.
This exposes a critical flaw in LLM-as-a-Judge benchmarks that treat single human annotations as ground truth.
Since individual judgments are unstable, an LLM aligning with them likely overfits to noise rather than capturing true perception.

\subsection{RQ2: Structural Analysis}

\begin{figure}
  \includegraphics[width=0.47\textwidth, trim=0cm 0.3cm 0cm 0.3cm, clip]{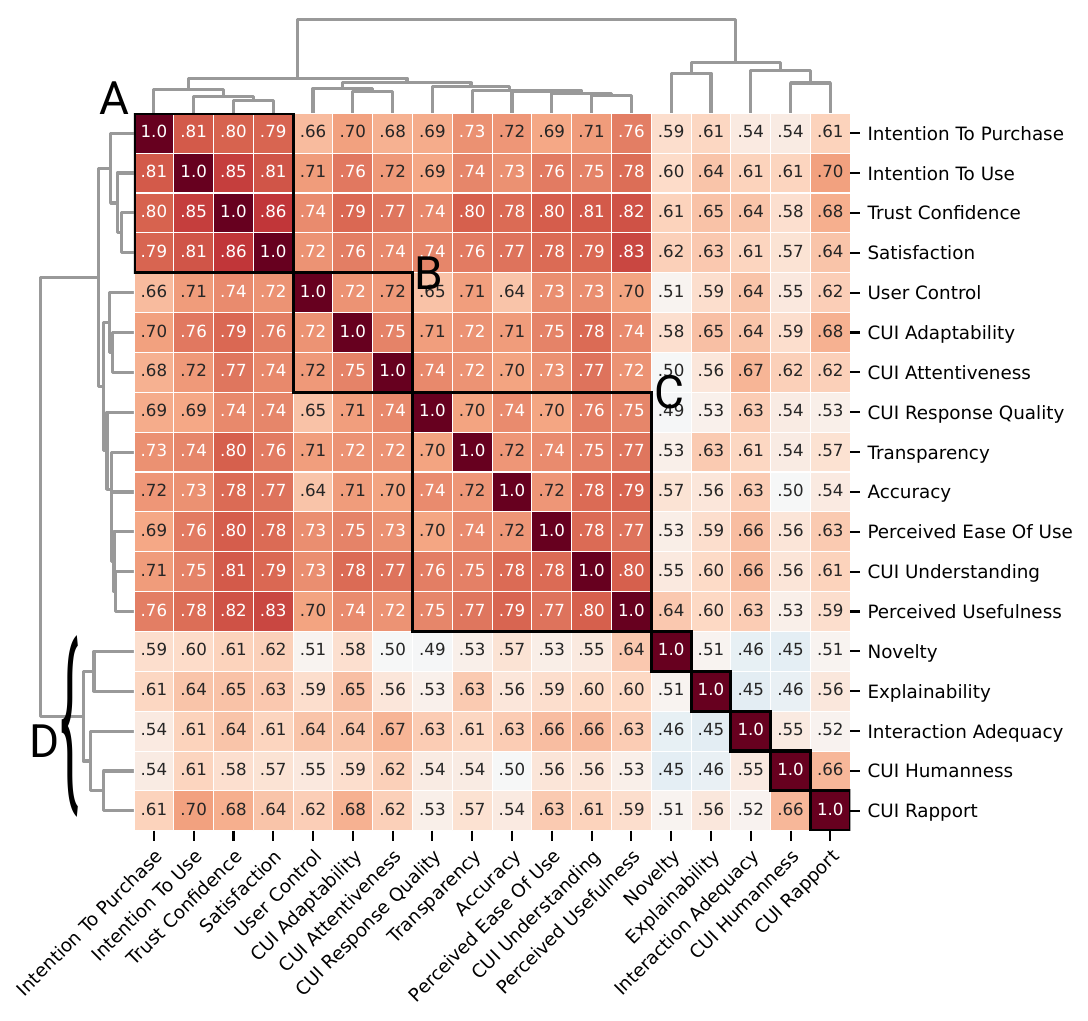}
  \Description{Heatmap displaying the pairwise Spearman correlations between different user-centric evaluation dimensions. The matrix is sorted via hierarchical clustering, showing distinct blocks of correlated dimensions.}
  \caption{Clustered correlation matrix of dimensions.}\label{fig:clustered_corr}
\end{figure}

To analyze dependencies between user-centric dimensions, we computed pairwise Spearman correlations over aggregated dialogue-level scores and applied hierarchical clustering to the resulting correlation matrix (Figure~\ref{fig:clustered_corr}).
Although the dendrogram yields eight clusters algorithmically, the salient structural split separates three highly correlated multi-item clusters (A--C) from a set of five largely independent singleton clusters, which we group as D for exposition, revealing substantial redundancy among CRS-Que dimensions in static evaluation.
Cluster~A comprises \emph{Satisfaction}, \emph{Trust/Confidence}, \emph{Intention to Use}, and \emph{Intention to Purchase}, forming the most tightly coupled block and reflecting a single holistic outcome judgment in which annotators collapse affective evaluation, trust, and behavioral intention into an overall assessment of system quality (the \emph{user satisfaction cluster}).
Cluster~B includes \emph{User Control}, \emph{CUI Adaptability}, and \emph{CUI Attentiveness}, capturing perceived responsiveness and user agency; these dimensions are strongly interrelated, empirically function as a single latent construct rather than distinct interaction properties.
Cluster~C consists of \emph{Accuracy}, \emph{Response Quality}, \emph{Understanding}, \emph{Ease of Use}, \emph{Usefulness}, and \emph{Transparency}, forming a coherent competence-oriented cluster in which annotators infer correctness, clarity, and usefulness from overlapping conversational cues and which occupies an intermediate position between interaction dynamics (Cluster~B) and evaluative outcomes (Cluster~A).
In contrast, \emph{Humanness}, \emph{Rapport}, \emph{Interaction Adequacy}, \emph{Novelty}, and \emph{Explainability} each form singleton clusters with substantially weaker correlations to the dominant quality signal; we group these dimensions as Group~D, as they capture socially grounded and experiential aspects of conversation that are not reducible to effectiveness or satisfaction.
In the movie domain, outcome, interaction, and competence dimensions collapse into a single global quality signal, revealing a pronounced \emph{halo effect} in third-party judgments.
While social and experiential aspects remain distinct but harder to assess, this suggests that offline evaluation can be simplified by aggregating highly correlated dimensions while selectively preserving socially grounded constructs.

\section{Limitations and Conclusion}

While our study provides valuable insights, it has limitations. First, using ReDial applies CUI-specific constructs to human-human transcripts.
Although human recommenders frequently serve as system proxies, static logs may not fully capture the dynamic nature of live human-AI interactions.
Second, ReDial conversations are inherently constrained and often lack complex dialog elements, such as detailed recommendation explanations.
Finally, our empirical scope is limited to a single dataset within the movie domain. 
Future research should validate these findings across diverse domains and interactive live-system evaluations.

Despite these limitations, our findings underscore a critical challenge: third-party annotators struggle to differentiate technical dimensions from general satisfaction in an offline setting.
The resulting halo effect causes ratings to collapse into a global quality signal, indicating that instruments like \mbox{CRS-Que} could be significantly simplified for static dialogue annotation.

Based on these insights, we derive four actionable recommendations for offline evaluation protocols:
(1) researchers should avoid relying on single-annotator scores, particularly for system-oriented constructs like \emph{Accuracy} or \emph{Transparency}, which are highly susceptible to noise;
(2) reliable measurement necessitates aggregating multi-rater feedback to mitigate individual biases;
(3) to combat the halo effect, evaluation designs should prioritize a small, stable core of dimensions (e.g., \emph{Satisfaction}, \emph{Perceived Usefulness}); and
(4) LLM-as-a-judge approaches should not be treated as absolute ground truth, but rather as single raters that require similar aggregation and validation strategies.

\begin{acks}
In accordance with ACM policy, we disclose the use of generative AI tools during the preparation of this manuscript.
Specifically, large language models were used for language refinement and code editing.
The authors have thoroughly reviewed, edited, and validated all generated text and code, and take full responsibility for the content, integrity, and originality of this paper.
\end{acks}

\bibliographystyle{ACM-Reference-Format}
\bibliography{references.bib}

\end{document}